\documentclass[11pt]{article}
\date{}
\usepackage{amsmath}
\usepackage{graphicx}
\def\b{\begin{equation}}
\def\e{\end{equation}}
\usepackage{color}
\usepackage{setspace}
\begin{document}
\title{\bf Quantum Hall effect and the different zero energy modes of graphene }
\author{M. R. Setare \thanks {Electronic mail: rezakord@ipm.com} , D. Jahani \thanks {Electronic mail: Dariush110@gmail.com}}
\maketitle {\it \centerline{  \emph{ Department of Science, Payame
Noor University, Bijar, Iran }} \maketitle {\it \centerline{
\emph{Young Researchers Club, Kermanshah Branch, Islamic Azad
University, Kermanshah, Iran. }}
\begin{abstract}
\emph{{The effect of an inhomogeneous magnetic field which varies
inversely as distance on the ground state energy level of
graphene is studied. In this work, we analytically show that
graphene under the influence of a magnetic field arising from a
straight long current-carrying wire ( proportional to the
magnetic field from carbon nanotubes and nanowires) exhibits zero
energy solutions. We find that contrary to the case of a uniform
magnetic field for which the zero energy modes show the
localization of electrons entirely on just one sublattice
corresponding to single valley Hamiltonian, zero energy solutions
in this case reveal that the probability for the electrons to be
on the both sublattices, say A and B, are the same.}}
\end{abstract}
\vspace{0.5cm} {\it \emph{Keywords}}: \emph{Graphene; Quantun Hall
effect; Zero energy modes. }

 \section {Introduction}
\emph{Graphene, a single layer of graphite, was isolated for the
first time in 2004 [1]. The carbons atoms in graphene are arranged
into a honeycomb structure which is consistent of the two
inequivalent triangular sublattices, say A and B [2]. Electrons in
graphene can hop to the nearest neighbours atoms which leads to the
formation of the two energy bands, each containing the same number
of states [3] and touching each other at the two inequivalent points
called Dirac points, say $\textbf{K}^{+}$ and $\textbf{K}^{-}$.
Around these points the energy dispersion relation of graphene is
linear in momentum which implies that it's low energy excitations
mimic the ultra relativistic massless particles. Thus, the low
energy excitations of graphene are described by the following
Dirac-like equation:
 \begin{equation}
 H=v_{F}\mbox{\boldmath$\sigma$.\textbf{p}}
 \end{equation}
 where $v_{F}$ is the Fermi velocity and
 \begin{boldmath}$\sigma$\end{boldmath}=$(\sigma_{x},\sigma_{y})$
 is the Pauli
matrices vector with $\sigma_{i}$, $i=x,y,z$, the $i$ Pauli matrix.
The above equation implies that the electrons in graphene behave as
massless charged Dirac fermions confined in a 2D space, an
interesting feature that real particles do not exhibit because all
the massless elementary particles happen to be electrically neutral.
These massless electrons shows peculiar properties which massive
relativistic carriers do not exhibit [4]. In fact the first
experimental evidence that revealed the charged carriers in graphene
mimic massless electrons was an unusual quantum Hall effect reported
in 2005 [5]. In spite of this fact that charge carriers in graphene
exhibit a four fold degeneracy (which comes from the real spin of
electrons in addition to another factor of two, due to the equal
contributions of the $\textbf{K}$-valleys, i.e. $\textbf{K}^{+}$ and
$\textbf{K}^{-}$) we see that in experiments instead of quantization
of the Hall conductivity in multiples of
\begin{equation}
\sigma_{xy}=4n\frac{e^2}{h}\ \ \ \ \ \ n\in
\{...,-2,-1,0,1,2,...\}
\end{equation}
it is observed that Hall conductivity is:
\begin{equation}
\sigma_{xy}=4(n+\frac{1}{2})\frac{e^2}{h}=4\nu\frac{e^2}{h},\ \ \
\ \ \ \nu=(n+\frac{1}{2}),
\end{equation}
which shows that the integer quantum Hall effect (IQHE), appears
in half-integers. It is also observed that unlike to the quantum
Hall effect for 2D conventional systems which appears in the
strong magnetic field and low temperature limit, the IQHE in
graphene can be observed even at the room temperature [6]. This is
because of ultra-relativistic nature of its charge carriers which
mimic the massless Dirac fermions. These massless charge carriers
as we'll show later, contrary to the conventional 2D systems, show
interesting results under the influence of a uniform magnetic
field. We, before discussing the effect of a constant magnetic
field $\textbf{B}=B_{z}$ perpendicular to the graphene's plane,
note that even for conventional 2D systems, the periodic potential
due to the host lattice is of no relevance to the quantum Hall
problem because the size of the electron wave packet in a magnetic
field is much larger than the lattice period. The periodic
potential due to the lattice is, therefore, neglected in studies
of the quantum Hall effect, however if one considers to calculate
the Landau levels based on the tight-binding model, the
commensurability problem between the magnetic flux and lattice
unit cell is needed to be considered. This problem is known to
inevitably occur in the two dimensional electron system [7-9].
However, interestingly for graphene the periodic potential of the
honeycomb lattice is already built-in and therefore it is counted
in the massless Dirac-like equation. Thus, we do not really need
to incorporate explicitly the periodic potential term into the
Dirac equation.\\
 Now, in order to obtain the energy spectrum of graphene in the presence of
 a uniform magnetic field which is considered to be perpendicular to
 the garphene's plane, by choosing the symmetric gauge and taking the units such that $c=1$, the
  single valley Hamiltonian of graphene
  can be written
 as:
 \begin{equation}
 H=v_{F}\left(%
\begin{array}{cc}
  0 & \Pi_{x}-i\Pi_{y} \\
 \Pi_{x}+i\Pi_{y} & 0 \\
\end{array}%
\right),
 \end{equation}
 where
\begin{equation}\mbox{\boldmath$\Pi$}=-i\hbar(\partial_{x}-\frac{iy}{2l^2},
\partial_{y}+\frac{ix}{2l^2}),\ \ \ \ \ l^2=\frac{\hbar }{B|e|}
\end{equation}
Then, one may write the equation (4) in the form of the following
eigenvalue equation:
 \begin{equation}
(i\hbar \partial_{t}+i\hbar v_{F}\sigma_{x} \Pi_{x} +i\hbar
v_{F}\sigma_{y} \Pi_{y})\Psi_{\textbf{K}^+}(\textbf{r},t)=0.
 \end{equation}
  Next, multiplying the above equation by $\sigma_{z}$ (the z-component of
Pauli matrix) gives:
\begin{equation}
( i\hbar\hat{\gamma}^0 \partial_{t}+i\hbar v_{F}\hat{\gamma}^1
\Pi_{x}+i\hbar v_{F}\hat{\gamma}^2
\Pi_{y})\Psi_{\textbf{K}^+}(\textbf{r},t)=0,
 \end{equation}
with the $\hat{\gamma}$ matrices as:
\begin{equation}
 \hat{\gamma}^0=\left(\begin{array}{cc}
1 & 0 \\
0 & -1 \end{array}\right)\ ,\
 \hat{\gamma}^1=\left(\begin{array}{cc}
0 & 1 \\
-1& 0 \end{array}\right)\ ,\
\hat{\gamma}^2=\left(\begin{array}{cc}
0 & -i\\
-i& 0\end{array}\right).
 \end{equation}
Here in order to solve the equation (7), we split the 2-spinor
$\Psi_{\textbf{K}^{+}}$ into its sublattice parts:
 \begin{equation}
\Psi_{\textbf{K}^{+}}(\textbf{r},t)=\left(\begin{array}{c}
\varphi  \\
 \psi\end{array}\right)e^{i\frac{E}{\hbar}t},
 \end{equation}
 which inserting it into the equation (7) leads
 us to the following expression:
\begin{equation}
\left(\begin{array}{c}
\  \ \dot{\varphi}  \\
-\dot{\psi}\end{array}\right)
 =\left(\begin{array}{cc}0 & S\\D &
 0\end{array}\right) \left(\begin{array}{c}\varphi
 \\\psi\end{array}\right),
 \end{equation}
where $S$ and $D$ are given by:
 \begin{equation}
 S=-v_{F}(\partial_{x}-\frac{iy}{2l^2})+iv_{F}(\partial_{y}+\frac{ix}{2l^2}),
 \end{equation}
 \begin{equation}
 D=v_{F}(\partial_{x}-\frac{iy}{2l^2})+iv_{F}(\partial_{y}+\frac{ix}{2l^2}).
 \end{equation}
There is no need to say that from equation (10) one can obtain the
following second order differential equations:
 \begin{align}
&\ddot{\varphi}=-SD\varphi, \\
 &\ddot{\psi}=-DS\psi,
 \end{align}
 which by introducing the following dimensionless
quantities:
\begin{align}
x\rightarrow \widetilde{x}=\frac{x}{l},\ \ \ \ \ y\rightarrow
\widetilde{y}=\frac{y}{l},
 \end{align}
 we can solve the them with
respect to $\varphi$ and $\psi$. In order to do so, we first need
to make the following ansatz:
\begin{align}
\varphi(\textbf{r},t)=e^{-iEt/\hbar}\varphi(\textbf{r}),
 \end{align}
 and
 \begin{align}
\psi(\textbf{r},t)=e^{-iEt/\hbar}\psi(\textbf{r}),
 \end{align}
 which plugging them in the equations (13) and (14) yields:
 \begin{align}
E^2=\frac{\hbar
^2v_{F}^2}{l^2}\left[(-i\partial_{\widetilde{x}}-\frac{\widetilde{y}}{2})^2+
(-i\partial_{\widetilde{y}}+\frac{\widetilde{x}}{2})^2\pm\frac{1}{l^2}\right].
 \end{align}
  where the positive (minus) sign corresponds to the solution for $\varphi$
  ($\psi$). Transforming to the complex coordinates:
\begin{equation}
 z=\widetilde{x}-i\widetilde{y},\ \ \ \bar{z}=\widetilde{x}+i\widetilde{y},
 \end{equation}
 gives the equation (18) as follows:
 \begin{align}
E^2=\frac{\hbar
^2v_{F}^2}{l^2}\left[-4\partial_{z}\partial_{\overline{z}}+\frac{1}{4}z\overline{z}+
\overline{z}\partial_{\overline{z}}-z\partial_{z}\pm1\right]
 \end{align}
 At this point,
 we can define the ladder operators $\hat{a}^{\dag}$ and
  $\hat{a}$ as:
 \begin{equation}
\hat{a}^{\dag}=\frac{1}{\sqrt{2}}(\frac{\bar{z}}{2}-2\partial_{z}),\
\ \
 \hat{a}=\frac{1}{\sqrt{2}}(\frac{z}{2}+2\partial_{\bar{z}}),
  \end{equation}
   which from them, one can write the solution for $\varphi$ in the following form:
  \begin{align}
E^2=\frac{2\hbar ^2v_{F}^2}{l^2}\left[\hat{a
}^{\dag}\hat{a}+1\right],
 \end{align}
 where $\hat{N}=\hat{a
}^{\dag}\hat{a}$, with the eigenvalues n=0,1,..., is the number
operator. While for $\psi$, we obtain the following solution:
   \begin{align}
E^2=\frac{2\hbar ^2v_{F}^2}{l^2}\hat{a }^{\dag}\hat{a}.
 \end{align}
  Here, the two solutions could be packed in one equation as:
 \begin{align}
E^2=\frac{\hbar ^2v_{F}^2}{l^2}\left[2(n+\frac{1}{2})\pm1\right]
 \end{align}
  At this point one can express the square of the Hamiltonian (4) in terms of the number operator $\hat{N}$:
\begin{equation}
 H^2=\frac{2\hbar^2 v_{F}^2}{l^2}\left(%
\begin{array}{cc}
  \hat{N}+1 &0\\
 0& \hat{N }\\
\end{array}%
\right),
 \end{equation}
 with the following eigenstates and eigenvalues for $H$:
 \begin{equation}
\Psi_{n,\textbf{K}^{+}}=\frac{1}{\sqrt{2}}\left(%
\begin{array}{c}
  |n-1\rangle \\
  \pm|n\rangle\\
\end{array}%
\right),\ \ \ \ E_{n}=\pm\frac{\hbar v_{F}}{l}\sqrt{2n}
 \end{equation}
It is clear that for $n\neq0$ we have one pairs of eigenstates and
eigenvalues but for $n=0$ (corresponding to $E=0$) we have:
  \begin{equation}
\Psi_{0,\textbf{K}^{+}}=\left(%
\begin{array}{c}
 0 \\
  |0\rangle\\
\end{array}%
\right)
 \end{equation}
 It is clear that the solution corresponding to the $\textbf{K}^{+}$ ($\textbf{K}^{-}$) point\footnote{The corresponding
eigenstate for $\textbf{K}^{-}$ is $\Psi_{0,\textbf{K}^{-}}=\left(%
\begin{array}{c}
 |0\rangle\\
  0\\
\end{array}%
\right)$  }
 shows that
 the probability for the electrons to be on the
 sublattice A (B) is zero. Thus, the zero solution corresponding
  to $\textbf{K}^{+}$ ($\textbf{K}^{-}$)
 implies the localization of Dirac fermions on the B (A) sublattice.
  In the original experimental paper it is argued that since in
all the $n\neq0$ energy levels both pseudospin states are filled,
whereas in the $n =0$ level only one is, the density of states in
the latter case is 1/2 that of the other levels and therefore it
contributes only $\frac{e^2}{2h}$ per spin/valley. \\The above
argument does not seem entirely satisfactory, since, as it is clear
from general solutions (26), any given eigenstate is normalized to
one irrespective of whether one pseudospin component is zero or not.
Hence the half contribution ($\frac{e^2}{2h}$ ) of the zero energy
mode corresponding to the single valley index could not be explained
in this way because, as we see from the normalized eigenstates,
electrons localize on just one sublattice, instead of being
contributed half on the sublattice A and half on the sublattice B.\\
There are another explanation for observation of half-integer
quantum Hall effect that says since for $n = 0$ we have only one
solution (as there is no difference between $+|0\rangle$ and
$-|0\rangle$) the degeneracy of this level is half of the other
energy levels for which there exist two solutions. What is wrong
about this conclusion is that existence of just one solution for
$n=0$ level (and two for others), does not simply mean that its
degeneracy is twice smaller because one of the two solutions
corresponds to the negative energy states (holes) and another to the
positive energy
states (electrons). Therefore this assumption could be disregarded.\\
 Another explanation might be based on this
assumption that the level $n=0$ is equally shared by electrons and
holes, meaning that it is half filled with electrons and half with
the holes, since there is no difference between $+|0\rangle$ and
$-|0\rangle$ in this level [10]. In the other words the ground state
energy level is completely filled with the same types of fermions
except the fact that they only differ by their charge which does not
prevent them from being subject to the Pauli exclusion principle. It
is by now that one can say the $n=0$ energy level contributes
$1/2(e^2/h)$. Hence, as for the other levels two kind of fermions
(holes and electrons) with the same number of states contribute in
the conductance, they contribute twice of $1/2(e^2/h)$ per
spin/valley.
 \\
The interesting feature that zero energy solutions exhibit,
motivate us to seek zero energy modes by examining the effect of
other types of magnetic fields on graphene's energy spectrum. In
fact, when the strength of the magnetic field is high the ground
state energy level is occupied by more and more electrons because
the degeneracy of the levels increase and therefore the lowest
levels play significant role in this case. In this paper, we
examine the effect of an inhomogeneous magnetic field which
varies as $B=(0,0,1/x)$ on the lowest energy level and show that
it exhibits zero energy solutions which is different from those
obtained for the case of the constant magnetic field discussed
above. As it is well-known, this magnetic field occurs around a
straight long current-carrying wire (see Fig. 1). We first, in
the next section briefly discuss the supersymmetric quantum
mechanics and the shape invariant method [11-14] which turns out
to be useful for our investigation.
\section{Supersymmetric quantum mechanics}
One of the methods for solving the quantum mechanical problems is
based on finding the relation between ground state wave function
and the corresponding potential. Considering the Hamiltonian
$H(x)$ as:
\begin{equation}
H(x)=-\frac{d^2}{dx^2}+V(x),
\end{equation}
with associated eigenfunctions and eigenvalues $\psi_{n}(x)$ and
$E_{n}$, respectively, we can write:
\begin{equation}
H(x)\psi_{n}(x)=E_{n}\psi_{n}(x)
\end{equation}
Now, if one defines $H_{1}(x)$ as:
\begin{equation}
H_{1}(x)\psi_{n}(x)=H(x)-E_{0},
\end{equation}
so that its ground state energy become zero ($E_{0}$ is the ground
state energy of $H(x)$), we can write:
\begin{equation}
H_{1}(x)=-\frac{d^2}{dx^2}+V_{1}(x)=-\frac{d^2}{dx^2}+V(x)-E_{0},
\end{equation}
It is clear that the two Hamiltonians $H(x)$ and $H_{1}(x)$ have
the same eigenfunctions. Denoting the eigenfunctions and
eigenvalues of $H_{1}(x)$ with $\psi_{n}^{1}(x)$ and $E^{1}_{n}$,
respectively, we can write:
\begin{equation}
H_{1}(x)\psi_{n}(x)=(E_{n}-E_{0})\psi_{n}(x)\ \ \rightarrow\ \
\psi_{n}(x)=\psi_{n}^{1}(x),\ \ \ E^{1}_{n}=E_{n}-E_{0}
\end{equation}
Now with defining the ladder operators $\hat{A}$ and
$\hat{A}^{\dag}$ as:
\begin{equation}
\hat{A}=\frac{d}{dx}+W(x),\ \ \ \ \
\hat{A}^{\dag}=-\frac{d}{dx}+W(x),
\end{equation}
where $W(x)$ is called superpotential, one can write $H_{1}(x)$ in
terms of the above operators as:
\begin{equation}
H_{1}(x)=\hat{A}^{\dag}\hat{A}=-\frac{d^2}{dx^2}+V_{1}(x),
\end{equation}
Here we see that from the relations (31) and (33) one can arrive
at the following relation for $V_{1}(x)$:
\begin{equation}
V_{1}(x)=W^2(x)-\frac{dW(x)}{dx}
\end{equation}
and keeping in mind that the ground state energy of $H_{1}(x)$ is
zero, we arrive at:
\begin{equation}
H_{1}(x)\psi_{0}^{1}(x)=\hat{A}^{\dag}\hat{A}\psi_{0}^{1}(x)=0,
\end{equation}
which means that $\hat{A}$ annihilates the ground state wave
function $\psi_{0}^{1}(x)$, i.e.:
\begin{equation}
\hat{A}\psi_{0}^{1}(x)=0
\end{equation}
Now it is obvious that from equations 34-36 one can write the
ground state wave function with respect to the superpotential
$W(x)$ and vise versa:
\begin{equation}
\psi_{0}(x)=Ne^{-\int^xW(x)dx}\ \ \leftrightarrow\ \
W(x)=-\frac{d}{dx}\ln
\psi_{0}(x)=-\frac{1}{\psi_{0}(x)}\frac{d\psi_{0}(x)}{dx}
\end{equation}
It is by now that we can define Hamiltonian $H_{2}(x)$, partner of
$H_{1}(x)$, which we denote them with $H_{+}(x)$ and $H_{-}(x)$
from now on, respectively, as follows:
\begin{equation}
H_{+}(x)=\hat{A}\hat{A}^{\dag}=-\frac{d^2}{dx^2}+V_{+}(x),
\end{equation}
with
\begin{equation}
V_{+}(x)=W^2(x,a_{0})+\frac{dW(x,a_{0})}{dx}
\end{equation}
The supersymmetric partner potential $V_{-}$ and $V_{+}$ are
supposed to be shape invariant if they satisfy the following
equation:
\begin{equation}
V_{+}(x,a_{0})=V_{-}(x,a_{1})+R(a_{0}),
\end{equation}
which means that two supersymmetric partner potentials have the
same form, but are characterized by the different values of
parameters $a_{0}$ and $a_{1}$ . To be more specific, the
parameter $a_{1}$ is a function of $a_{0}$, namely, $a_{1}$ =
R($a_{0}$) with R an independent function of $x$. Now one can
obtain the energy spectrum associated to $V_{-}(x, a_{0})$ simply
from the shape invariance condition as:
\begin{equation}
E_{n}^{-}(a_{0})=\sum_{i=0}^{n-1}R(a_{i}), \end{equation} where
for $n>0$: \b E_{0}^{-}(a_{0})=0\e while the corresponding wave
functions are given by:
\begin{align} \psi^{-}_{n}(x,a_{0})&\sim
\hat{A}^{\dag}(x,a_{0})\hat{A}^{\dag}(x,a_{1})...\hat{A}^{\dag}
(x,a_{0})(x,a_{1})\psi^{-}_{0}(x,a_{n})\nonumber\\
&=\hat{A}^{\dag} (x,a_{0})\psi^{-}_{n-1}(x,a_{1}).
\end{align}
Note that there are only a few problems that satisfy the shape
invariant condition (41). As we show in the next section, although
the ground state energy level can be obtained analytically, the
shape invariant condition is not satisfied.
\section{Zero energy modes corresponding to effect of a varying magnetic field}
 As we'll show in what follows, graphene spectrum under the
influence of a magnetic field which varies as inverse of distance,
i.e. $\textbf{B}=(0,0,1/x)$ exhibits zero energy modes. This
magnetic field occurs often, as it is the magnetic filed around a
long, straight current-carrying wire. In fact because of the
symmetry of the wire the magnetic lines are circles concentric
with it and lie in the planes perpendicular to the wire. The
magnetic field $\textbf{B}$ is constant on any circle of radius
$R$ and is given by:
\begin{align}
\textbf{B}=\frac{\mu_{0}I}{2\pi R}
\end{align}
where $I$ is the current of the wire and $\mu_{0}$ is the
magnetic constant. Now if we consider a graphene sheet which lies
parallel to the axis of wire so that the lines of the magnetic
field intersect the graphene sheet which is assumed to be in
xy-plane, the corresponding vector potential can be written as:
 \b\textbf{A}=(0,q\ln x,0),\ \ \rightarrow
\textbf{B}=(0,0,q\frac{1}{x}),\e where we have used the Landau
gauge and defined q to be:
\begin{align}
q=\frac{\mu_{0}I}{2\pi }.
\end{align}
At this point, if we go through the same procedure as the case of
the constant magnetic field (see section 1), in this case, we'll
obtain for the $S$ and $D$ (with taking $v_{F}=1$ in our
evaluations):
\begin{align}
S&=-\partial_{x}+i(\partial_{y}-iqe\ln x),\nonumber\\
D&=\partial_{x}+i(\partial_{y}-iqe\ln x).
\end{align}
 In the next step by taking the units such that $\hbar=c=1$, we can make the following ansatz:
\begin{equation}
\Psi(\textbf{r},t)=\left(%
\begin{array}{c}
  \varphi \\
  \psi\\
\end{array}%
\right)e^{iEt},
\end{equation}
which leads us to the following equation:
\b
E^2\psi(\textbf{r})=\left\{-\mbox{\boldmath$\nabla$}^2+q^2e^2\ln^2x+2iqe\ln
x\partial_{y}+\frac{qe}{x}\right\}\psi(\textbf{r}).\e
We then make the following ansatz for $\psi(\textbf{r})$ as:
\b\psi(\textbf{r})=e^{ik_{y}y}f(x),\e
which plugging it into (50) gives:
 \b \left[-\frac{d^2}{dx^2}+k^2_{y}+q^2e^2\ln^2x-2qek_{y}\ln
x+\frac{qe}{x}\right]f(x)= E^2f(x)\e
The above equation is an eigenvalue equation that can be written
as: \b H(x)f(x)= \epsilon f(x),\ \ \ \ \epsilon=E^2\e
  It is by now that we can write the
superpotential $W(x)$ in the form: \b W(x)=-a\ln x+\frac{b}{a}.
 \e
 Then we define Hamiltonian $H_{1}(x)$ as:
 \b H_{1}(x)=-\frac{d^2}{dx^2}+V_{1}(x)=-\frac{d^2}{dx^2}+V(x)-\epsilon_{0},\e
 where $\epsilon_{0}$ is the ground state energy of $H(x)$ and $V_{1}(x)$
by the use of the relation:
  \b V_{1}(x)=W^2(x)-\frac{d W(x)}{dx},\e
 is given by:
\b V_{1}(x)=a^2\ln^2x-2b\ln x+\frac{a}{x}+\frac{b^2}{a^2}.\e Now
by comparing the two Hamiltonians $H(x)$ and $H_{1}(x)$, one can
get $a$ and $b$ as:
 \b a=qe, \ \ \ b=qek_{y}, \e
  which reveals that they are just the same and, therefore, we obtain: \b
\epsilon_{0}=0, \e meaning that the ground state energy level,
$E_{0}$, is zero. Here we should note that the other energy levels
can not be derived analytically because the shape invariant
condition (41) is not satisfied. We also note that for $\varphi(r)$
the same result is obtained, since we the commutation relation:
\begin{equation}
[S,D]=0,
\end{equation}
  is satisfied, meaning that the $x$ and $y$ components of dynamical momentum do not commute with each other.
   Thus, we have arrived at a very important result. For
graphene under a varying magnetic field discussed above, there
exists two zero energy modes for which the probability for Dirac
fermions to be on the sublattice $A$ is the same as sublattice $B$.
However this does not mean that these zero energy are different to
those obtained for uniform magnetic field in the sense of living
electrons on the two sublattices. It is because, as we pointed out
in the first section, the wave function for the other Dirac point
are swapped for a constant magnetic field and therefore in an
uniform magnetic field electrons are present on both triangular
sublattices as well as the electrons in a varying magnetic field.
The whole argue is about the single valley Hamiltonian which One can
also obtained $f_{0}(x)$ as follows:
 \b f_{0}(x)=N\exp(eqx\ln x-eqx-k_{y}x)\e
where $N$ is the normalization constant and subscribe $0$ shows that
$f_{0}(x)$ is the wave function associated to the lowest energy
level. Note that, here $x$ takes its positive values ($x>0$). It may
be assumed that the existence of the two Dirac points has completely
fixed the zero of the energy at these points, however,
interestingly, for a magnetic field which varies inversely as square
of distance, i.e. $\textbf{B}$=($0,0,\lambda/x^2$) (where $\lambda$
is a constant), the energy spectrum can be obtained analytically
using the shape invariant method as [15]:
 \b E_{n}=\pm v_{F}\sqrt{k_{y}^2-\frac{\lambda^2 e^2k_{y}^2}{(n+1/2+\sqrt{1/4+\lambda e(\lambda e-1)})^2}},\e
 which shows that for $k_{y}\neq 0$, the $n=0$ energy level ($E_{0}$) is not zero, unlike the ground state energy
 due to the effect of the magnetic field from a long current-carrying wire which
 revealed (even for nonzero values for
$k_{y}$ ) to be zero.
\section{Implications for experiment}
No zero energy modes are observed when a magnetic field is applied
to a system consistent of electrons confined in a conventional two
dimensional structure, since charge carriers in conventional 2D
systems obey the schr\"{o}dinger equation of motion and therefore
no massless carriers are imagined for them. However, it does not
mean that strong magnetic field can not be applied to these
systems which give rise to the observation of conventional
quantum Hall effect. The problem is about the magnetic field
discussed in the previous section. In fact, one arrives at no
analytical solution when the effect of the magnetic field
$\textbf{B}=(0,0,q/x)$ (see Eq. (46)) is examined on the massive
carries no mater whether they behave relativistically or not.
This may be the reason why no investigations has been reported up
to now concerning the effect that this kind of magnetic field
might have on conventional 2D system. \\ From the above discution
we see that graphene could be considered as the only 2D structure
that investigation regarding the effect of the magnetic field
(46) - both from the theoretical and experimental point of view -
is worth noting.
  As it is shown in Fig.1, the magnetic lines lie in the
planes perpendicular to the wire and intersect the graphene's plane.
The magnetic field B which is constant on any circle of radius R,
decrease inversely as the distance increases in the
$x$-direction.\\Here, there is no need to say that the result
reported in this paper regarding the existence of zero energy modes
could be put to the test in contrast with other types of nonuniform
magnetic fields such as that varies inversely as square of the
distance $x$ (see equation 62) and those investigated in [16].
  \begin{figure}
 \begin{center}
\includegraphics[width=6cm]{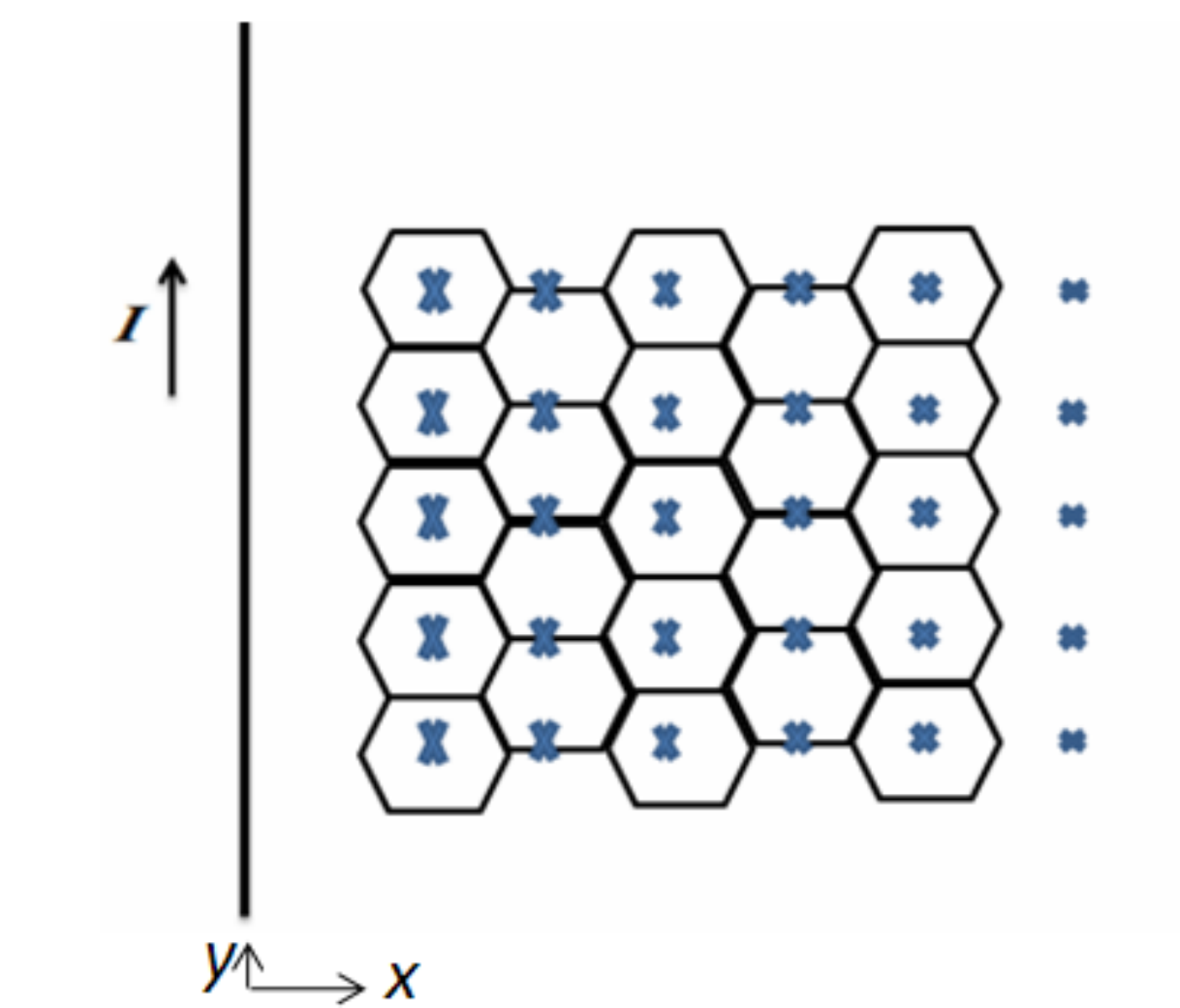}
 \hspace{0.1cm}
 \includegraphics[width=6cm]{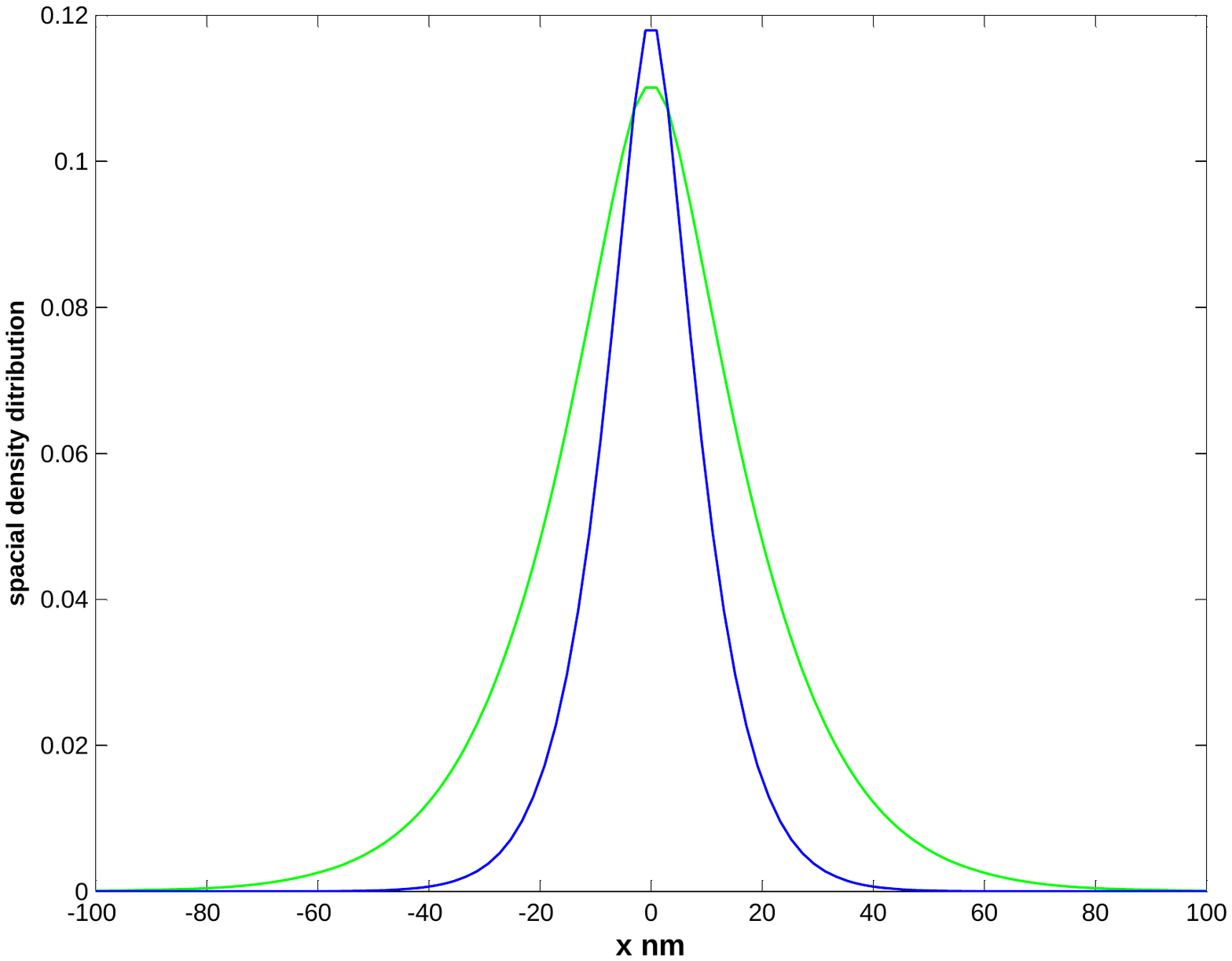}
\caption{Left: The lines of the magnetic field from a current
carrying wire which in the xy-plane decreases as $1/x$, lie in the
planes perpendicular to the wire and intersect the graphene's
sheet. Right: The density distribution associated with the
$f_{0}(x)$ for currents $I_{1}$ (green line) and $I_{2}$ (red
line) for which $I_{2}=2I_{2}$ . As it is expected the width of
the wave-packet grows as the current I -and therefore the
magnetic field- decreases.}
\end{center}
   \end{figure}
\\  In the end of this section, we should note that the
magnetic field created by a toroid for a special case varies
inversely as distance as well. However, it is often used to create
an almost uniform magnetic field in some enclosed area. One can use
Ampere's law to obtain the magnetic field inside of a toroid with N
turns of wire as:
\begin{equation}
 B=\frac{\mu
NI}{2\pi r}
\end{equation}
 where r which is measured from the center of the toroid is the radius of a circle to which the direction of the
 magnetic field is tangent.
In fact the magnetic filed is approximately uniform inside the
torus, if the radius of toroid, r, is very large compared with the
cross-sectional radius of it. But for small values of r the magnetic
field falls off inversely as r. So our results can also hold for
this case as well.
\section{Conclusion}
In this work, we examined the effect of a magnetic field varying
inversely as distance on the ground state energy level of
graphene. One important reason for studying this type of magnetic
field- apart from this fact that it occurs often-is that it is .
In fact it is the magnetic field of a long carrying-current wire
and, therefore, it can be important when it comes to applications
of carbon nanotubes and nanowires with graphene. We also showed
that graphene under the influence of such a magnetic field
exhibits zero energy modes which is kind of different from the
zero energy modes corresponding to the uniform magnetic filed
(counted for the observation of the unconventional quantum Hall
effect in graphene). In fact, contrary to the former case, the
zero energy solutions associated to the magnetic field
$B=(0,0,q\frac{1}{x})$ do not show the localization of Dirac
fermions on just one sublattice but they imply that the
probability to find electrons on one sublattice, say A, is the
same as other one, say B. We also discussed the original
interpretation of observation of the half-integer quantum Hall
effect in graphene which does not seem to be complectly
satisfactory because the localization of electrons on one
sublattice does not imply that the density of states due to the
$n=0$ Landau level is half of the others. We also discussed about
how the effect of the two kind of magnetic field which varied as
$1/x^2$ and $1/x$ on the graphene spectrum could lead to the
different results. \\In this work we investigated the effect of a
the latter case on the massless Dirac fermions of undoped
graphene, leading to observation of two zero energy modes which,
as we said, are different in the sense of living the electrons on
the different sublattices (per valley/spin). As we pointed out,
considering the massive relativistic particles no analytical
solution for the lowest energy level is obtained and it might be
the reason that the potential (45) have not been considered up to
now.\\In the end, we should note that at the first sight it might
seem strange that the localization of charge carriers differs for
the varying and constant magnetic field. However, by considering
the two Dirac points we see that one indicates the localization
of electrons on B and another on the A sublattice and therefore
the equivalency of carbon atoms is not broken. \\Another point
which is worth noting here is that  the magnetic energy levels
obtained from the tight-binding model agrees well with that
calculated from the kp model \cite{h17}. }

\end{document}